\RequirePackage{ifpdf}
\ifpdf 
\documentclass[pdftex]{sigma}
\else
\documentclass{sigma}
\fi

\begin{document}

\renewcommand{\PaperNumber}{073}

\FirstPageHeading

\ShortArticleName{Do All Integrable Evolution Equations Have the Painlev\'e Property?}

\ArticleName{Do All Integrable Evolution Equations\\ Have the Painlev\'e Property?}

\Author{K.M. TAMIZHMANI~$^\dag$, Basil GRAMMATICOS~$^\ddag$ and Alfred RAMANI~$^\S$}

\AuthorNameForHeading{K.M. Tamizhmani,  B. Grammaticos and A. Ramani}

\Address{$^\dag$~Departement of Mathematics, Pondicherry University,
Kalapet, 605014 Puducherry, India} 
\EmailD{\href{mailto:tamizh@yahoo.com}{tamizh@yahoo.com}}

\Address{$^\ddag$~IMNC, Universit\'e Paris VII-Paris XI, CNRS, UMR
8165, B\^at. 104, 91406 Orsay, France}
\EmailD{\href{mailto:grammaticos@univ-paris-diderot.fr}{grammaticos@univ-paris-diderot.fr}}

\Address{$^\S$~Centre de Physique Th\'eorique, Ecole Polytechnique,
CNRS, 91128 Palaiseau, France}
\EmailD{\href{mailto:ramani@cpht.polytechnique.fr}{ramani@cpht.polytechnique.fr}} 

\ArticleDates{Received June 12, 2007; Published online June 19, 2007}

\Abstract{We examine whether the Painlev\'e property is necessary for
the integrability of partial dif\/ferential equations (PDEs). We show
that in analogy to what happens in the case of ordinary dif\/ferential
equations (ODEs) there exists a class of PDEs, integrable through
linearisation, which do not possess the Painlev\'e property. The same
question is addressed in a discrete setting where we show that there
exist linearisable lattice equations which do not possess the
singularity conf\/inement property (again in analogy to the
one-dimensional case).}


\Classification{34A99; 35A21; 39A12}

\noindent
The Painlev\'e property has been quintessential in the domain of
integrable systems. While the modern era of integrability was spurred
by the derivation (or rediscovery in some cases \cite{[1]}) of
evolution equations  integrable by spectral methods (S-integrable systems
in the Calogero terminology \cite{[2]}), the Painlev\'e approach has
almost from the outset become an unmatched integrability test leading
to the discovery of a slew of new integrable systems. The conjecture~\cite{[3]},
formulated by Ablowitz and Segur (in collaboration with one of us,
A.R.) and which came to be known as the ARS conjecture, states that:
``Every ordinary dif\/ferential equation which arises as a reduction of
a completely integrable partial dif\/ferential equation is of Painlev\'e
type (perhaps after a transformation of variables)''. This conjecture
was soon improved thanks to Weiss and collaborators who managed to
treat partial dif\/ferential equations directly, without the constraint
of considering reductions  \cite{[4]}.

We can illustrate the situation through the classical example of the
modif\/ied-KdV equation:
\begin{gather*}
u_t=u_{xxx}-6u^2u_x.
\end{gather*}
We expand the solution around a singularity manifold $\phi(x,t)$ 
(the calculations are very simple so there is no need to use the simplifying Kruskal ansatz~\cite{[5]}). 
We f\/ind the expansion
\[
u=\pm\big(\phi_x/\phi-\phi_{xx}\phi_x^{-1}/2
+((\phi_{xxx}-\phi_t)\phi_x^{-2}/6-\phi_{xx}^2\phi_x^{-3}/4)\phi
+c_2\phi^2+c_3\phi^3+\cdots\big)
\]
 which is
obviously singlevalued, and possesses the full complement of integration
constants, satisfying thus the Painlev\'e property.

The discrete analogue of the Painlev\'e property is singularity
conf\/inement. As conjectured \cite{[6]} by two of the authors (B.G and A.R.) in
collaboration with V.~Papageorgiou ``any singularity spontaneously
appearing in an integrable discrete system must disappear after a few
iteration steps''. (The fact that singularity conf\/inement does not
suf\/f\/ice as a discrete integrability criterion and must be complemented
by the f\/initeness of the Nevanlinna order \cite{[7]} of the solution has been
discovered in \cite{[8]} and was amply commented in several publications \cite{[9],[10]}).
The
discrete analogue of the mKdV equation
\begin{gather*}
x_{m+1,n+1}=x_{m,n}{x_{m+1,n}-cx_{m,n+1}\over
cx_{m+1,n}-x_{m,n+1}},
\end{gather*}
where $c$ is a constant, has a singularity whenever, due to specif\/ic
initial conditions, $x$ vanishes or becomes inf\/inite. However this
singularity disappears in the next iteration step and does not
propagate at all: it is immediately conf\/ined.

An impressive number of examples have by now established the fact that
both partial dif\/ferential and dif\/ference equations integrable through
spectral methods possess the Painlev\'e property, (or equivalently, have
conf\/ined singularities). These results can be extended in 
a straightforward way  to the case of ordinary equations (both  dif\/ferential
  and dif\/ference ones, this last family encompassing $q$-dif\/ference
equations as well). This is a quite natural result since many ordinary
equations can be obtained as reductions of partial ones.  However it is
in the domain of ordinary equations that exceptions to the reciprocal of
the above statement have been obtained. Indeed, we found in \cite{[11]} that there
exist equations which are integrable through linearisation and which do
{\sl not}  possess the Painlev\'e property. This is true both in the
continuous  and the discrete case. We shall illustrate this through two
examples.

We start from   the linear equation
\begin{gather}\label{eq3}
{t x''+(at-1/2) x'+bt x\over  x''+a x'+b x}=K
\end{gather}
and take its derivative so as to eliminate $K$, obtaining a third order
equation. Next we show  that the same third order equation can be obtained
if we start  from the nonlinear equation
\begin{gather}\label{eq4}
x''x'+2ax'^2+3bx'x+(2ab-b')x^2=M
\end{gather}
and take its derivative so as to eliminate $M$. Here $a$ and $b$ are not
free. We have
$b=a^2-a'/2$ and $a$ satisfying the equation
\begin{gather*}
a'''=6a''a+7a'^2-16a'a^2+4a^4
\end{gather*}
which is equation XII in the  Chazy classif\/ication \cite{[12]}.
So, equation \eqref{eq4} is integrable by linearisation
through equation \eqref{eq3}. It is straightforward to show that \eqref{eq4} violates the
Painlev\'e property. Solving it for $x''$, we f\/ind  terms
proportional to $x^2/x'$ (and $1/x'$) which were shown to be incompatible
with  the Painlev\'e property \cite{[13]}.

A {\sl caveat} is in order at this point. While there  exist large classes
 of linearisable equations without the Painlev\'e property, there does
also exist a family of  linearisable equations  which do satisfy the
Painlev\'e criterion. The best known example of equations belonging  to
this class are the Riccati equation and its higher order analogues \cite{[14]}.

We turn now to an example of a discrete equation and examine the mapping
\begin{gather}\label{eq6}
x_{n+1}=ax_{n-1}{x_n-a\over x_n-1}.
\end{gather}
As shown in \cite{[15]}, two   singularities exist   when either   $x=1$ or   $x=a$.
The f\/irst singularity is conf\/ined leading to a f\/inite singularity
pattern $\{ 1,\infty , a\}$.  The second  singularity never conf\/ines
unless $a=1$ (in which case the mapping is trivial) or   $a$ is a cubic
root of unity (with  the resulting mapping being periodic with period six).

On the other hand, \eqref{eq6} is linearisable. Indeed, introducing
$y_n=x_nx_{n-1}-x_n-ax_{n-1}$
we  reduce \eqref{eq6} to the linear mapping $y_{n+1} = ay_n$.

Of course the  remark concerning the existence of linearisable systems
with the Painlev\'e property has its analogue here for linearisable
mappings with conf\/ined singularities: all mappings of the ``projective''
family fall into this class~\cite{[16]}.

It is thus natural given the results presented above on PDE's integrable
through  spectral methods which possess the Painlev\'e property, and
linearisable ODE's  without it, to wonder whether linearisable PDE's
without the Painlev\'e property may exist.
Calogero has stressed the importance  of the existence of PDE's integrable
by methods dif\/ferent from the spectral ones. He has dubbed the members of
this class C-integrable systems. So, quite understandably, we sought
linearisable PDE's  among the lengthy list of C-integrable systems
established by Calogero  \cite{[17]}. We shall not go into a fully detailed analysis
of the equations of the Calogero classif\/ication, but present just  the
outcome of our investigation. It turned out that the C-integrable
equations presented by  Calogero  can be shown to  belong to one of two
classes.

The f\/irst class comprises equations which  can be reduced to  a linear one
thanks to some transformation involving the integral of the variable of
the nonlinear  equation. A typical example is:
\begin{gather}\label{eq8}
u_t-u_{xxx}=3u_{xx}u^2+9u_{x}^2u+3u_{x}u^4
\end{gather}
that is linearised to
\begin{gather*}
v_t-v_{xxx}=0
\end{gather*}
through
\begin{gather}\label{eq10}
v(x,t)=u(x,t)\exp\int^x u(x',t)^2 dx'.
\end{gather}
Of course, the fact that the integral of $u^2$  appears in the
transformation is not creating problems. As a matter of fact, it is
possible to rewrite  \eqref{eq10} as
\begin{gather*}
{v_x \over v}={u_x \over u}+u^2={w_x \over 2w}+w
\end{gather*}
which is just a Riccati equation for $w=u^2$. Quite expectedly, equation
\eqref{eq8} (and others, similar to it)  possess the  Painlev\'e property.

The second class of the C-integrable systems of Calogero  comprises
equations which  are obtained from some  other integrable (sometimes
linear)  equations  through hodograph  transformations. The prototypical
equation of this class is the Dym equation \cite{[2]}
\begin{gather*}
u_t=u^3u_{xxx} 
\end{gather*}
which is  related to the KdV equation. Equations of this class quite often
 possess the `weak' Painlev\'e property. This  is the case for the Dym
equation. The  expansion around a singularity manifold $\phi(x,t)$ is
\[
u_0\phi(x,t)^{2/3}+\sum\limits_{p=1}^\infty u_p\phi(x,t)^{(p+2)/3}.
\]
Moreover, there exist equations in the Calogero list of C-integrable PDE's
belonging to the class of solvable through   hodograph  transformations
which do not satisfy the Painlev\'e property at all.
An example  of such an equation is $u_t=f(u_x)/u_{xx}+g(u_x)+uh(u_x)$,
where $f$, $g$ and $h$  are arbitrary functions.
However    due to the very  special nature of the hodograph
transformation we consider that such examples do not constitute a
satisfactory answer to our quest.

So the question remains:  do linearisable PDE's  without the Painlev\'e
property exist? The answer to this question is an unqualif\/ied ``yes".
Let us construct  a specif\/ic example. We shall adopt the construction we
follow for the derivation of the Burgers' equation. For the latter, we
start from a linear equation $v_t+v_{xx}=0$ and obtain  a nonlinear  one
through a Cole--Hopf relation  $v_x+uv=0$.  In order to derive the equation
we are seeking, we start from a nonlinear, linearisable (Riccati) equation
in one variable $v_t+v^2=0$
and couple it through a Cole--Hopf like relation to  another variable in a
new direction $u_x+uv=0$. Eliminating $v$ we obtain a nonlinear equation
for $u$:
 \begin{gather}\label{eq13}
 uu_{xt}-u_xu_t-u_x^2=0.
 \end{gather}
 This is obviously a linearisable equation since its solution proceeds
through  the solution of one linearisable equation  and a linear one, in
cascade. The  solution of this equation does {\sl not} possess  the
Painlev\'e property.
Instead of performing a standard  Painlev\'e  analysis  let us prof\/it
from the fact that the solution of \eqref{eq13}  can be explicitly constructed.
Solving the equation for $v$ we f\/ind $v=(t-\phi(x))^{-1}$. Next we
integrate for $u$ and obtain $\log u=-\int(t-\phi(x))^{-1}dx$. A
singularity will appear  in the expansion of $u$ whenever we have $x=\xi$
such that $\phi(\xi)=t$. We solve for $\xi$ and f\/ind $\xi=\psi(t)$  (where
 $\psi$ is the inverse function of $\phi$).
Expanding   $\phi(x)$ around $\xi$  we have
$\phi(x)=\phi(\xi)+(x-\xi)\phi'(\xi)+\cdots$ and the integration for $u$
can be performed order by order. We f\/ind $u\propto
(x-\psi(t))^{\psi'(t)}+\cdots$. Thus, since the exponent $\psi'(t)\equiv
1/\phi'(\psi(t))$  is arbitrary, the solution does {\sl not} possess the
Painlev\'e property.

Equation \eqref{eq13} may be easily generalised. The  principle remains   the
same. One starts  from a~linearisable equation in  one independent and one
dependent variable, say $v(t)$.  If, for instance, we take for $v$  a
higher-order projective equation, we are guaranteed  that the solution
for $v$ will satisfy the Painlev\'e property.
 Next we couple this equation to a linear  PDE  of the form
$f(v)u_x+g(v)u_t+h(v)u=0$, where $f$, $g$,  and $h$  can be taken  as
inhomogeneous linear  functions of $v$. Eliminating $v$ one obtains an
equation for $u$ which  is linearisable  and can be shown to violate the
Painlev\'e property, the exponent of the leading singular term being
again an arbitrary function of $t$.

 We now  turn to the case of a lattice  linearisable equation.  We start
with a simple homographic mapping (the index $m$ is dummy at this level)
\begin{gather*}
v_{m,n+1}+1+{1\over v_{m,n}}=0
\end{gather*}
 and couple it to  a linear equation
\begin{gather*}
u_{m+1,n}-u_{m,n}v_{m,n}=0.
\end{gather*}
  Eliminating $v$ we f\/ind for $u$ the equation
\begin{gather}\label{eq16}
u_{m+1,n+1}u_{m+1,n}+u_{m,n+1}u_{m+1,n}+u_{m,n}u_{m,n+1}=0.
\end{gather}
While this equation is linearisable, it does {\sl not} have conf\/ined
singularities. Indeed, if at some lattice position  we have $u_{m,n}=0$
(which is perfectly possible given the appropriate  initial conditions)
iterating \eqref{eq16} we f\/ind that $u_{k,n}=0$ for all $k\ge m$. On the other
hand, since~\eqref{eq16}  is linearisable, we expect the growth  of the sequence
of its iterates  to be linear. This turns out to be indeed the case.
Taking  initial conditions $u_{0,0}={\rm const}$, $u_{0,n}=a(n)+b(n)p/q$,
$u_{m,0}=c(m)+f(m)r/s$ (with $a$, $b$, $c$, $f$ arbitrary functions of their
argument) and computing  the global homogeneous degree  $d_{m,n}$ in
$p$, $q$, $r$, $s$, we f\/ind that $d_{m,n}=m+2$   for $m>0$.

 Generalising \eqref{eq16} is quite straightforward. It suf\/f\/ices to start from a
linearisable equation for $v$ of higher order  (of which several
examples do exist). Next we couple $v$ to a linear equation of the form
  $f(v_{m,n})u_{m+1,n}+g(v_{m,n})u_{m ,n+1}+h(v_{m,n})u_{m,n}=0$ where
$f$, $g$, $h$ are  f\/irst degree  in $v$, and using the f\/irst equation we eliminate
$v$. We surmise that the equation for $u$ will in general have
unconf\/ined singularities. However this has to be  examined on a per
case basis  since there does not seem to exist a general argument for
the singularity structure of the f\/inal equation.

The results we presented  above can be further generalised. As a matter of
fact, we do not have  to take for the f\/irst equation one that  satisf\/ies
the Painlev\'e property, or possesses conf\/ined singularities. It suf\/f\/ices
that  it be linearisable  for our construction to go through (although
here the Painlev\'e property, continuous or discrete,  would be already
violated at the f\/irst step).
For the continuous case, we start from a linearisable equation in one dimension
\begin{gather}\label{eq17}
v_{tt}=(v_t-1)/v
\end{gather}
which is linearisable through the system $v_t=yv+1$, $y_t+y^2=0$.
The general solution of \eqref{eq17}
is $v=(c_1(x)+\log (t+c_2(x))(t+c_2(x))$, which clearly violates the Painlev\'e property. 
(A singular solution for \eqref{eq17} also exists: $v=t+c_2(x)$).
Next we couple the equation for $v$ to an equation for $u$  through $u_x-uv=0$ and obtain
\begin{gather*}
u_{xtt}=u_{xt}\left(2{u_t\over u}+{u\over u_x}\right)+{u_{tt}u_x\over u}
-2{u_xu_t^2\over u^2}-u_t-{u^2\over u_x}
\end{gather*}
which is linearisable without having the Painlev\'e property.
Turning to the discrete case we consider the one-dimensional  mapping (again, the index $m$ is dummy at this level)
\begin{gather}\label{eq19}
y_{m,n+1}=y_{m,n}+y_{m,n}^2/y_{m,n-1}
\end{gather}
which is obtained from the linearisable system $z_{n+1}=z_n+1$ and $y_{n+1}=z_ny_n$.
The solution of~\eqref{eq19} is $A(m)\Gamma(n+c(m))\equiv B(m)(-1)^n/\Gamma (1-n-c(m))$ with
$ \pi A(m)=B(m)\sin (c(m)\pi)$. This solution is regular unless $c(m)$ is an integer. 
If $c(m)=N$, an integer, the f\/irst expression, with nonzero $A$, has poles on a 
half-inf\/inite line $n\le-N$, and is regular for $n>-N$. 
The second expression, for f\/inite $B$, is regular for 
$n\le-N$ and zero for  $n>-N$ . In both cases we have a~typical nonconf\/ined singularity. 
Coupling equation \eqref{eq19} with  $x_{m+1,n}-y_{m,n}x_{m,n}=0$ leads to
\begin{gather*}
x_{m+1,n+1}x_{m+1,n-1}x_{m,n}^2=x_{m+1,n-1}x_{m+1,n}x_{m,n+1}x_{m,n}+x_{m+1,n}^2x_{m,n+1}x_{m,n-1}.
\end{gather*}
Again, we have a linearisable equation with unconf\/ined singularities.
Many more examples can be explicitly constructed  underlining the
fact  that there exist  linearisable PDE's which do not satisfy the
Painlev\'e property.

\pdfbookmark[1]{References}{ref}
\LastPageEnding

\end{document}